\documentclass[runningheads]{llncs}
\usepackage{graphicx}
\usepackage{multirow}
\usepackage{lscape}
\usepackage{hyperref}
\usepackage{enumitem}
\usepackage{amssymb}
\begin{document}
\title{Object-Centric Event Logs: Specifications, Comparative Analysis and Refinement} 
\titlerunning{OC Event Logs: Specifications, Comparative Analysis and Refinement }
%
\author{Alexandre Goossens\inst{1}\orcidID{0000-0001-8907-330X},
Johannes De Smedt\inst{1} \orcidID{0000-0003-0389-0275} \and
Jan Vanthienen\inst{1}\orcidID{0000-0002-3867-7055}}
%
%
\authorrunning{A. Goossens et al.}

\institute{Leuven Institute for Research on Information Systems (LIRIS), KU Leuven \email{\{FirstName\}.\{LastName\}@kuleuven.be}}
\maketitle              
\begin{abstract}
Process mining aims to comprehend and enhance business processes by analyzing event logs. 
Recently, object-centric process mining has gained traction by considering multiple objects interacting with each other in a process. This object-centric approach offers advantages over traditional methods by avoiding dimension reduction issues. However, in contrast to traditional process mining where a standard event log format was quickly agreed upon with XES providing a common platform for further research and industry, various object-centric logging formats have been proposed, each addressing specific challenges such as object relations or dynamic attribute changes. This makes that interoperability of object-centric algorithms remains a challenge, hindering reproducibility and generalizability in research. Additionally, the object-centric process storage paradigm aligns well with a wide range of object-oriented databases storing process data.

This paper introduces a specifications framework from three perspectives originating from process mining (what should be analyzed), object-centric process modeling (how it should be modeled), and database storage (how it should be stored) perspectives in order to compare and evaluate object-centric log formats. By identifying commonalities and discrepancies among these formats, the study delves into unresolved issues and proposes potential solutions. Ultimately, this research contributes to advancing object-centric process mining by facilitating a deeper understanding of event log formats and promoting consistency and compatibility across methodologies.
\end{abstract}
\newpage
%
%
%
\section{Introduction}

The general aim of process mining is to understand, discover and enhance business processes by analyzing event logs recording the different steps occurring in a process.
Traditionally, these processes are analyzed through the lens of a single object such as a customer or a product.
However, not all processes necessarily fit in this single point of view paradigm as was already indicated in 2001 with the introduction of proclets to model a process from multiple perspectives\cite{van2001proclets}.
The idea that processes can be analyzed from multiple angles has recently gained interest and is more commonly known as object-centric process mining \cite{berti2023advancements}. 
Compared to traditional process mining, object-centric process mining does not suffer from typical dimension reduction issues such as convergence issues (incorrect multiplication of activities which do not match their actual occurrence) and divergence issues (ambiguity in which activities belong to which objects).

Since the introduction of object-centric process mining, various object-centric logging formats have been proposed with each of them addressing relevant, but different, object-centric process issues such as the eXtensible Object-Centric (XOC) logs \cite{van2017object} and Object-Centric Behavioral Constraint (OCBC) models \cite{li2017automatic} addressing the need of object relations in object-centric processes as well as the Artefact-Centric Event Log (ACEL) proposal \cite{m2023process}.
However, these proposals had scalability issues which were addressed with the Object-Centric Event log (OCEL) 1.0 format \cite{ghahfarokhi2020ocel} at the expense of storing object relations.
The Data-aware Object-Centric Event Log (DOCEL) format in turn addressed the issue of changing attribute values in a process.
This latter issue was also addressed with the recent OCEL 2.0 format which also includes object relations \cite{berti2023ocel}.
Next to that, there is a proposal to store object-centric processes in so-called Event Knowledge Graphs which come with their own set of challenges \cite{fahland2022process}.
All these proposals are happening whilst the IEEE-group is also proposing their own standard object-centric process event log format called Object-Centric Event Data (OCED)\footnote{https://www.tf-pm.org/resources/oced-standard} which is used as a guideline by the different object-centric event log proposals. 
Even though there have been other proposals such as MXML \cite{van2005meta} prior to the introduction of the eXtensible Event Stream (XES) log format \cite{xes2014ieee}, the field of process mining greatly benefited from having the XES standard as it provided a common ground to conduct further research \cite{augusto2018automated,van2022process}. 
Compared to the XES standard, however, there is currently no consensus on how to store object-centric processes in logs. 
This suggest that while there are different object-centric process mining algorithms these do not necessarily support other object-centric event log formats limiting the reproducability and generalizability of object-centric process mining research such as in  \cite{goossens2024extracting,adams2022oc}.
Hence, the object-centric process mining field would benefit from a study investigating what the different object-centric event log formats share and do not share with one another and in doing so hopefully find common ground to have an 'Object-Centric XES'.

This study aims to address this gap by analyzing the different object-centric event log formats using a framework providing an overview of the different issues that an object-centric event log format could or should address. 
The object-centric specifications framework is built by investigating three different fields crucial to support a process mining exercise: storage given that a wide range of databases are inherently object-centric instead of process-centric making the extraction of object-centric event logs more natural from such databases \cite{li2018extracting}, extraction to discover and analyze the stored processes, and representation.
Next, a thorough discussion explores solutions for unresolved issues not agreed upon in various object-centric event log formats or left unaddressed altogether.

The remainder of the paper is structured as follows: Section \ref{Framework construction}  outlines the framework's specifications and its derivation. 
Next, using the constructed framework, Section \ref{Discussion} compares the different object-centric event log formats and on how they do or do not address the different issues followed by a lengthy discussion on the insights from this comparison in Section \ref{Insights from the Comparison}.
Next, Section \ref{Object-Centric Event Log Refinement} proposes further refinements for object-centric event logs.
Finally, Section \ref{Limitations and Future Work} discusses the limitations and future work of this study with Section \ref{Conclusion} concluding this paper.

\section{Framework Dimensions and Specifications} \label{Framework construction}

In the next section, object-centric event log format specifications are derived by investigating three relevant research fields for object-centric processes namely:
\begin{enumerate}
    \item Process Mining: Given that object-centric process mining is a continuation of process mining we start from its initial conception and investigate this field for specifications for object-centric event logs.
    \item Conceptual Modeling: Given that object-centric process mining aims at mining object-centric process models and/or object models, it is necessary to investigate how these models are structured and how objects are intertwined with processes in this field.
    \item Databases: Object-centric event log formats are inherently linked to the storage of data. Therefore investigating how different types of databases such as object-oriented databases or temporal databases deal with this is relevant.
\end{enumerate}
For each research field, the most relevant issues are identified and their implications for object-centric processes are discussed. 

\subsubsection{Framework Dimensions}
Object-centric process mining merges process mining and conceptual modelling, focusing on how multiple object instances interact during process execution. It integrates the \textbf{Event-to-Event (E2E) Dimension} from traditional process mining, which studies event sequences' relationships, and the \textbf{Object-to-Object (O2O) Dimension} from conceptual modeling, which explores object relationships. Additionally, it considers the interplay between events and objects, known as the \textbf{E2O} dimension. These dimensions are pivotal in understanding object-centric analysis and storage. A recent study highlights these dimensions' significance, suggesting they can manifest individually or concurrently in an analysis or event log \cite{berti2023advancements}. Apart from \textbf{E2E, O2O and E2O}, one other dimension emerges from the analysis: the \textbf{Data Quality Dimension}, offering specifications on data quality.


\subsection{Process Mining Research Field Perspective}

To gather specifications from the process mining perspective, two primary sources are identified: the process mining manifesto \cite{van2012process} and the latest object-centric process mining overview study \cite{berti2023advancements}.

\subsubsection{Process Manifesto: Process mining types}The process mining manifesto outlines three types of process mining: process discovery, conformance checking, and enhancement (prediction and prescription).
It is clear that any object-centric event log format must support these three types of process mining as well.

\subsubsection{Process Manifesto: Process Log Perspectives}Additionally, the process mining manifesto identifies the following perspectives a traditional process log should have:

\begin{enumerate}
    \item Control-flow: It is important to know how all the activities are related to one another.
    \item Organizational perspective: Who does what in an organization?
    \item Data-perspective: Associate data with events. Enhance the log with accurate information for each event.
    \item Time-perspective: Assign timestamps to events for accurate time-based analysis of processes.
\end{enumerate}
These perspectives need to be transformed to fit into an object-centric process mining context.

\subsubsection{Process Manifesto: Event Data Criteria} The process mining manifesto also identifies the following criteria for event data:
\begin{enumerate}
    \item Trustworthy: The information stored in the event log can be trusted i.e. activities are not duplicated and all cases contain their respective activities.
    \item Complete: There is no crucial information missing due to a transformation i.e. convergence or divergence issues.
    \item Well-defined semantics: No ambiguity exists in the data format or attributes.
    \item Safe: information is anonymized.
\end{enumerate}
While all preceding criteria for event data are crucial, only the first three are impacted by the event log format itself. Ensuring data integrity is primarily the responsibility of data engineers rather than the format alone.

\subsubsection{Object-Centric Process Mining} 
The main finding of \cite{berti2023advancements} is that no identified log format enables analysis on all object-centric dimensions (\textbf{E2E, O2O, and E2O}); instead, some formats support two out of three possibilities \cite{berti2023advancements}.
The study did not aim to deeply compare the differences between event log formats.
Since then, new formats addressing all three dimensions (\textbf{E2E, O2O, and E2O}) have emerged, such as OCEL 2.0 and Event Knowledge Graphs (EKGs) \cite{berti2023ocel,fahland2022process}.

Additionally, the study \cite{berti2023advancements} categorizes different object-centric process mining algorithms into these three analysis dimensions. It is not unexpected that most algorithms utilize event logs supporting \textbf{E2O and O2O} dimensions, as this aligns with the primary objectives of object-centric process analysis and the availability of these logs initially. Intriguingly, the algorithms predominantly yield process models emphasizing \textbf{E2E}, while the other two dimensions (\textbf{E2O and O2O}) are less frequently discovered.

\subsubsection{Resulting Framework Specifications}
From process mining research, it is essential to note that an object-centric event log format must comply with the requirements outlined in the process mining manifesto. The following \textbf{E2E} specifications are derived from this:
\begin{itemize}
        \item \textit{An event must have a unique event ID} 
        \item \textit{An event must have a timestamp} 
        \item \textit{An event must have an activity}  
        \item \textit{An event can have other optional attributes} 
\end{itemize}
These specifications correspond to those in the XES format \cite{xes2014ieee} and the process mining manifesto \cite{van2012process}. Furthermore, the \textbf{O2O} specifications are derived from the \textbf{E2E} specifications adhering to the same principles as events in the XES format.
\begin{itemize}
     \item \textit{An object must have an object type} 
     \item \textit{An object must have a unique object ID } 
     \item \textit{An object can have optional object attributes}
\end{itemize} 


\subsection{Object-centric Process Modeling Perspective}

Object-centric conceptual modeling integrated with processes is predominantly found in two domains, each approaching the issue from a distinct perspective. On one side, the research field of data-aware process modeling treats processes as objects interacting with one another, characterized by well-defined behavior \cite{kunzle2013object,iDOCEM,snoeck2014enterprise}. Conversely, process mining begins with a process model where objects interact with one another. 
Both will be analyzed to identify common points to derive key specifications.
\begin{enumerate}[noitemsep]
    \item \textbf{Object relations: } In data-aware process modeling, an explicit object model defining object relations is defined prior to modeling the object behavior in a process \cite{kunzle2011philharmonicflows,kunzle2013object,snoeck2014enterprise}. In object-centric process modeling, such an object model is not necessarily explicitly defined like with proclets \cite{van2001proclets} or Colored Petri Nets (CPN) \cite{kleijn2012regions,ghilardi2022petri}. A recent Petri Net-based approach models the process of multiple individual object instances for read-only data \cite{van2022data}. Conversely, OCBC models explicitly define an object model alongside a declarative process model \cite{van2017object}, similar to Guard-Stage Milestone (GSM) \cite{hull2011business}. Generally, within the Unified Modeling Language (UML), object relations precede modeling UML behavior diagrams \cite{UML2017}. The idea behind all these approaches is that objects and their relations impact a business process. Storing object relations within an event log can reveal insights into process execution. For instance, why a customer may only have one order at a time or why multiple orders can be packaged together for delivery.    
    \item \textbf{Object-Event relations: } In all mentioned approaches, objects are linked to events either through individual object behavior models resembling state machines \cite{UML2017,kunzle2011philharmonicflows,kunzle2013object,snoeck2014enterprise} or directly integrated into the process model \cite{van2001proclets,kleijn2012regions,ghilardi2022petri,van2022data,van2017object}. Regardless of how objects and events are connected, it is essential for both to be associated with each other. This information should also be stored in event logs; otherwise, obtaining a complete view of an object-centric process becomes impossible, as knowing when and which objects interact in a process is crucial. Notably, some approaches require at least two events per object (one for creation and one for deletion), such as PhilHarmonic Flows \cite{kunzle2011philharmonicflows}, Merode \cite{snoeck2014enterprise}, and UML state machine diagrams \cite{UML2017}. Consequently, every event in these approaches must relate to at least one object. These assumptions have significant implications; no object is considered relevant for a process if it is not involved in an event, and likewise, no event is deemed part of a process without involving at least one object. This means that system-wide events must impact all objects, rather than none. Certain object-centric event log formats incorporate this requirement like OCEL 1.0 \cite{ghahfarokhi2020ocel} and DOCEL \cite{goossens2023enhancing}.    
    \item \textbf{Object information can be updated: } In data-aware process modeling, events can update objects \cite{kunzle2011philharmonicflows,snoeck2014enterprise}, a concept also found in UML sequence and activity diagrams. Although the approach in \cite{van2022data} primarily focuses on read-only immutable data, it acknowledges the existence of processes involving mutable data. Similarly, in process mining, the XES standard permits updating case information. This principle should be extended to object-centric processes, as a comprehensive view of object evolution can elucidate routing decisions or final outcomes within a process.
    \item \textbf{Individual behavior of an object must be recoverable: } Both data-aware process modeling and object-centric process modeling approaches aim to understand how individual object instances undergo a process to analyze exceptions or sub-processes.  Additionally, the XES standard explicitly supports this analysis by linking every process execution to an individual trace. Therefore, for a comprehensive understanding of how multiple objects undergo a process simultaneously, an object-centric event log format must allow the discovery of the individual sequence of events involving each object.
\end{enumerate}

\subsubsection{Resulting Framework Specifications}
Based on this analysis, the following specifications for the \textbf{O2O} dimensions are added:
\begin{itemize}
    \item \textit{Object attributes can change values and those changes are traceable} 
    \item \textit{An object instance can be related to other object instances} 
    \item \textit{Supports relation qualifiers}
\end{itemize}
The first specification addresses the need to track changes in object information over time within a process, as processes often interact with both read-only and read-write data.
The last specification emphasizes the importance of capturing object relationships, a fundamental aspect of conceptual modeling where objects are interconnected. 
This idea also comes back in object-centric process modeling where object relations are also important.
The last specification proposes including relation qualifiers between events and objects or between two objects, endorsed by studies in the object-centric process field \cite{berti2023advancements}, OCEL 2.0 \cite{berti2023ocel}, and OCED.

For the \textbf{E2O} dimension, the following specifications are derived:
\begin{itemize}
    \item \textit{An event must be related to 1..N objects}
    \item \textit{An object must be related to 1..N events}
    \item \textit{The process of an individual object instance must be recoverable} 
\end{itemize}
The first two specifications indicate the minimum cardinalities there are between events and objects, with a stricter formulation chosen for its significant impact on what can be stored in an object-centric event log.
The last specification ensures the recoverability of the complete process of a single object instance, aligning with concepts proposed in PhilHarmonic Flows \cite{kunzle2011philharmonicflows} and the process mining manifesto's definition of a trace \cite{van2012process}.

\subsection{Databases Perspective}
The databases research field is relevant as it addresses storage issues directly within databases, insights from which may be applicable to object-centric event logs.
This section primarily discusses object-oriented databases, as they share similarities with object-centric process storage \cite{bertino1991object,atkinson1990object}. Additionally, it touches upon temporal \cite{snodgrass1986temporal} and moving databases \cite{wolfson1998moving}, as they handle changing information over time, a crucial aspect of process mining.
Knowledge graphs \cite{hogan2021knowledge,9416312} are also considered, as certain proposals have utilized them as a foundation for storing object-centric processes \cite{fahland2022process}.

Regarding object-oriented databases, the following relevant requirements are found from the object-oriented databases manifesto \cite{atkinson1990object}: 
\begin{enumerate}[noitemsep]
    \item \textbf{Complex objects: }  The manifesto emphasizes the need for databases to support complex objects and attributes such as lists, sets, and tuples, reflecting the complexity often found in objects. This notion also applies to object-centric processes.
    \item \textbf{Object Identity:} It is a requirement for object-oriented databases that each object instance must have a unique object ID otherwise it is difficult to uniquely trace the object instance over time \cite{atkinson1990object,bertino1991object}.
    \item \textbf{Object types and classes: } In conceptual modeling, objects can be categorized by type or class, each with its own set of attributes or operations \cite{atkinson1990object,bertino1991object}. 
    \item \textbf{Inheritance:} Object-oriented databases incorporate the concept of inheritance, where an object type inherits attributes and behavior from another object type, including multiple inheritance. However, this aspect has not yet been explored in object-centric process mining.
    \item \textbf{Updating Object Information:} Object-oriented databases may create new object versions for each update, necessitating a direct link to the previous object version \cite{bertino1991object,atkinson1990object}. Temporal databases, introduced in 1986, distinguish between transaction time and valid time \cite{snodgrass1986temporal}. Transaction time denotes when a change occurred in the database, while valid time reflects the time when the change occurred in reality and is accurate \cite{PELEKIS_THEODOULIDIS_KOPANAKIS_THEODORIDIS_2004,snodgrass1986temporal}.
    An object-centric event log should only contain accurate and unambiguous information, implying it should only include valid time information.
    The interplay between transaction time and valid time should be left for the databases themselves. Lastly, moving databases \cite{wolfson1998moving} address storing objects with rapidly changing values, distinguishing between static and dynamic attributes based on how frequently their values change over time.
    The variety of database types addressing this issue highlight the importance of accurately tracking changing attributes values over time in an object-centric process.
    \item \textbf{Schema Evolution:} In \cite{stonebraker1990third}, the significance of supporting changes in a database schema for third-generation databases was highlighted, also mentioned as optional in the object-oriented database manifesto \cite{atkinson1990object}. This observation can extend to object-centric processes, where the object model itself may change during a process, potentially affecting the process execution, similar to process mining's identification of concept drift. In object-centric process mining, an \textit{object drift} could induce concept drift later on.  It might therefore be interesting to investigate how to support this in object-centric event log formats as well.
    \item \textbf{Knowledge Graphs: } Whilst not discussed in the object-oriented databases manifesto \cite{atkinson1990object}, knowledge graphs store entities as nodes, connected by relations \cite{hogan2021knowledge,9416312}. 
    They typically use tuples in the form (tail, relation, head, sometimes (timestamp)), where tail and head are nodes.
    Knowledge graphs are highly extensible, accommodating various node types. Consequently, they can support updates in object information, complex objects, and object type inheritance. In an object-centric context, this could involve distinct node types for object types, events, and attributes. Therefore, knowledge graphs seem to be able to accommodate for almost the above requirements.
\end{enumerate}

The object-oriented database manifesto outlines additional mandatory features such as \textbf{computational completeness}, which ensures any computational function can be expressed in the database language, and \textbf{database recovery}. While crucial for databases, these features are irrelevant for an object-centric event log format.

\subsubsection{Resulting Framework Specifications}

From the databases research field, there is a strong interest in the \textbf{O2O} dimension confirming the importance of the object-object specifications \cite{atkinson1990object}.
It mainly confirms that it is considered good practice to store object information in that manner. 
Since object-oriented databases also support object inheritance, further exploration of this concept in object-centric process mining could be valuable. The following \textbf{O2O} dimensions are found from the database research field:

\begin{itemize}
    \item \textit{Object relations can change over time}
    \item \textit{Supports object type inheritance}
 \end{itemize}   

 The first specification, highlighted in database studies \cite{stonebraker1990third} and data-aware process modeling \cite{kunzle2011philharmonicflows,snoeck2014enterprise}, is exclusively featured in the ACEL object-centric event log format proposal \cite{m2023process}. The last specification, allowing for object type inheritance, was underscored in database proposals, considering its close relation to data models \cite{atkinson1990object,bertino1991object}, though currently, none of the object-centric event log formats supports this specification.

Next, we identify the following specifications in the \textbf{Data Quality} Dimension, drawn from the process mining manifesto and the databases research field:
\begin{itemize}
    \item \textit{Data changes must be unambiguous}
    \item \textit{Data is uniquely identifiable}
    \item \textit{Data should be minimally duplicated}
    \item \textit{Data storage should be maximally scalable}  
\end{itemize}
The four data quality specifications address crucial aspects of data storage: accuracy, uniqueness, consistency, and scalability. An object-centric event log format should enable traceable value changes, ensure unique identification of objects and events over time, minimize data duplication, and ensure scalability of the event log itself.

\section{Comparison of the different Object-Centric Event Log Formats} \label{Discussion}



A framework for object-centric event log specifications is developed from the previous analysis, consisting of four dimensions: \textbf{E2E, O2O, E2O and Data Quality}. The first three dimensions align with the classification of object-centric process techniques proposed in \cite{berti2023advancements}, while the \textbf{Data Quality Dimension} is crucial considering the event log's role in data storage. 
These specifications represent common insights from different research fields, each with its specific focus on object-centric processes or storage, and should be viewed as results rather than requirements.

Table \ref{Specifications} compares various object-centric event log formats on the specifications. 
A \textit{\checkmark} indicates support for the specification, while a \textit{X} indicates lack of support, with clarifications provided if necessary.
The comparison includes only object-centric event log formats and the OCED standard, encompassing:

\begin{itemize}
    \item \textbf{XOC\cite{li2018extracting}}: XOC stores both object-centric data and the complete object model alongside the event that triggered a change in the object model.
    \item \textbf{OCEL 1.0\cite{ghahfarokhi2020ocel}}: OCEL 1.0 was developed after XOC to address scalability issues. It consists of an events table and an objects table. However, it does not support dynamic attributes (attributes changing values over time) \cite{goossens2023enhancing} or object-object relations \cite{m2023process}.
    \item \textbf{ACEL\cite{m2023process}}: ACEL was developed after OCEL 1.0 and mainly addresses the need to store object-object relationships within an object-centric process. 
    \item \textbf{DOCEL\cite{goossens2023enhancing}}: DOCEL was developed after OCEL 1.0 and supports dynamic attributes with the use of foreign keys.
    \item \textbf{OCEL 2.0\cite{berti2023ocel}}: The successor of OCEL 1.0 incorporates dynamic attributes using timestamps and includes object relations in the objects table.
    \item \textbf{EKG\cite{fahland2022process}}: Based on knowledge graphs, this storage formats creates node types for events, attributes, timestamps and object types. These are then connected to one another with the use of relations.
    \item \textbf{OCED}: It is essential to note that OCED is currently a metamodel, not an object-centric event log format. However, it is included in the comparison because it is developed by the IEEE group responsible for XES, and several event log formats, such as OCEL 2.0 and EKG drew inspiration from OCED.
\end{itemize}

The comparison therefore does not include the XES format \cite{xes2014ieee} (which covers all the \textbf{E2E} specifications but not the \textbf{O2O or E2O} dimensions) or other interesting event log formats which solve specific issues such as IoT \cite{bertrand2023nice}.

\begin{table}[]
\caption{Specifications of object-centric Event Log Formats}
\label{Specifications}
\resizebox{\textwidth}{!}{%
\begin{tabular}{|l|l|l|l|l|l|l|l|}
\hline
\textbf{Specifications} & \textbf{XOC} \cite{li2018extracting} & \textbf{OCEL 1.0} \cite{ghahfarokhi2020ocel} & \textbf{OCEL 2.0} \cite{berti2023ocel} & \textbf{DOCEL} \cite{goossens2023enhancing} & \textbf{ACEL} \cite{m2023process} & \textbf{EKG}\cite{fahland2022process} & \textbf{OCED} \\ \hline
\textbf{E2E} &  &  &  &  &  &  &  \\ \hline
S1: An event must have a unique event ID & \checkmark & \checkmark & \checkmark & \checkmark & \checkmark & \checkmark & \checkmark \\ \hline
S2: An event must have a timestamp & \checkmark & \checkmark & \checkmark & \checkmark & \checkmark & \checkmark & \checkmark \\ \hline
S3: An event must have an activity & \checkmark & \checkmark & \checkmark & \checkmark & \checkmark & \checkmark & \checkmark \\ \hline
S4: An event can have other optional attributes & \checkmark & \checkmark & \checkmark & \checkmark & \checkmark & \checkmark & \checkmark \\ \hline
\textbf{O2O} &  &  &  &  &  &  &  \\ \hline
S5: An object must have object type & \checkmark & \checkmark & \checkmark & \checkmark & \checkmark & \checkmark & \checkmark \\ \hline
S6: An object must have a unique objectID & \checkmark & \checkmark & \checkmark & \checkmark & \checkmark & \checkmark & \checkmark \\ \hline
S7: An object can have optional object attributes & \checkmark & \checkmark & \checkmark & \checkmark & \checkmark & \checkmark & \checkmark \\ \hline
\begin{tabular}[c]{@{}l@{}}S8: An object can change values\\  and those are traceable\end{tabular} & \begin{tabular}[c]{@{}l@{}}X\\ (not traceable)\end{tabular} & \begin{tabular}[c]{@{}l@{}}X\\ (not traceable)\end{tabular} & \checkmark & \checkmark & \checkmark & \checkmark & \checkmark \\ \hline
\begin{tabular}[c]{@{}l@{}}S9: An object instance can be\\  related to the other object instances\end{tabular} & \checkmark & X & \checkmark & X & \checkmark & \checkmark & \checkmark \\ \hline
S10: Supports relation qualifiers & X & X & \checkmark & X & X & X & \checkmark \\ \hline
\begin{tabular}[c]{@{}l@{}}S11: Object relations can change\\ over time (schema evolution)\end{tabular} & X & X & X & X & \checkmark & X & X \\ \hline
S12: Support object type inheritance & X & X & X & X & X & X & X \\ \hline
\textbf{E2O} &  &  &  &  &  &  &  \\ \hline
S13: An event must be related to 1..N objects & X & \checkmark & X & \checkmark & X & X & X \\ \hline
S14: An object must be related to 1..N events & X & \checkmark & X & \checkmark & X & X & X \\ \hline
\begin{tabular}[c]{@{}l@{}}S15: The process of an individual\\  object instance must be recoverable\end{tabular} & \checkmark & \checkmark & \checkmark & \checkmark & \checkmark & \checkmark & \checkmark \\ \hline
\textbf{Data Quality Dimension} &  &  &  &  &  &  &  \\ \hline
S16: Data changes must be unambiguous & \checkmark & X & X & \checkmark & \checkmark & \checkmark & / \\ \hline
S17: Data is uniquely identifieable & \checkmark & X & X & \checkmark & \checkmark & \checkmark & / \\ \hline
S18: Data should be minimally duplicated & X & \checkmark & \checkmark & \checkmark & \checkmark & \checkmark & / \\ \hline
S19: Data storage should be maximally scalable & X & \checkmark & \checkmark & \checkmark & \checkmark & \checkmark & / \\ \hline
\end{tabular}%
}
\end{table}
Regarding the \textbf{E2E} dimension, all the object-centric event logs formats agree on the 4 specifications.
For the \textbf{O2O} dimension, the different formats all incorporate specifications \textit{S5 to S7}. 
For specification \textit{S8}, the idea is that an object attribute can be changed and that it can be traced back. 
A key-point here is that tracking attribute changes can be done with timestamps (OCEL 2.0), foreign keys (DOCEL or ACEL), or even object versions (XOC and OCEL 1.0) but in case object versions are used it is important to know which object was the previous version which neither XOC or OCEL 1.0 support.
The next \textbf{O2O} specification (\textit{S9}) deals with storing the object relations an object instance has. 
This specification is not supported by OCEL 1.0 and DOCEL.
Regarding specification \textit{S10}, only the OCEL 2.0 supports relation qualifiers as prescribed by OCED.
For \textit{S11}, ACEL is the only standard supporting schema evolution whilst no event log format supports object type inheritance (\textit{S12}).

Regarding the \textbf{E2O} dimension, the first two specifications (\textit{S13 and S14}) deal with a more constraint view of what can be stored in an object-centric event log namely only events with an object or only objects with an event. 
This idea is only required by two formats namely OCEL 1.0 and DOCEL. 
The other formats allow also events and objects to be stored which are not related to objects and events respectively to be stored.
For \textit{S15}, all event log formats allow to recover the control-flow of a specific object instance.

Moving to the \textbf{Data Quality} dimension, \textit{S16} states that data changes must be unambiguous however for OCEL 1.0 and OCEL 2.0 this is not guaranteed.
Given that in OCEL 1.0 attribute changes have to be stored with the events, it is unclear to which object instance these changes apply \cite{goossens2023enhancing}.
For OCEL 2.0, attribute value changes are given a timestamp but in case two events happen simultaneously using the same object instance it is unclear which event is responsible for this value change.
This is also why OCEL 1.0 and OCEL 2.0 do not follow specification \textit{S17}.
For \textit{S18}, only XOC does not minimally duplicate information given that for each event that updates an object relation, the whole object model is returned which in turn makes the event log format not very scalable and hence does not support specification \textit{S19} neither \cite{ghahfarokhi2020ocel}.
Regarding \textit{S19}, two additional event log formats need to be discussed. 
OCEL 2.0 allows to store object relations but given that object models can be very large this might get very quickly very large.
The same observation goes for ACEL where changing attribute values and relationship changes are stored together with the events themselves which might also not scale very well. 
From the previous observations, it is clear that storing object-object relationships is a challenge when it comes to scalability but it comes with the advantage of providing a more holistic overview of an object-centric process.

\section{Insights from the Comparison} \label{Insights from the Comparison}

\subsubsection{Attribute value Changes}A first interesting insight is the fact that different object-centric event log formats allow for attribute value changes but that they differ in their implementation:
\begin{itemize}
    \item \textbf{OCEL 2.0: } As explained in the previous section, OCEL 2.0 allows to link a timestamp to an attribute value. This makes it possible to know when an attribute value was changed. However, it does not guarantee that the event responsible for this change can be known. In case an object instance is involved in two events at the same time, it is not possible to know which event is responsible for the attribute value change.
    \item \textbf{DOCEL: } In order to store changing object attribute values, DOCEL stores these in a different table where it makes use of two foreign keys namely the object ID to indicate to which object instance the changed value belongs to and the event ID to indicate which event is responsible for the change.
    \item \textbf{ACEL: } In order to store changing attribute values, ACEL stores a subtable containing the objectID, attribute and the new value, together in the event row responsible for the change. Even though this solves the issue, ACEL also stores in a similar manner the changing object-relationships making the overall event table not adhering to the normalization principles limiting a scalable and easy analysis \cite{10.7551/mitpress/12274.003.0034}. 
    \item \textbf{EKG: } For EKGs it is possible to link an attribute value to an event or a timestamp. Both are possible.
\end{itemize}

Regarding attribute value changes it seems that all the possible ways to store these have been covered (including using object versions by XOC and OCEL 1.0). 
Whether a certain way of storing these dynamic attributes is better mainly depends on what exactly is the purpose of the final analysis. In case an analysis focuses on both control-flow and the dynamic attributes, it is important to know that certain event log formats do not allow to analyze dynamic attribute changes. 

\subsubsection{Object Relations}Moving on to object-object relationships, it is clear from XOC that storing complete object models with each event quickly causes scalability issues \cite{goossens2023enhancing,ghahfarokhi2020ocel}.
However even with OCEL 2.0 and ACEL which also support the object-object relations, scalability can not necessarily be guaranteed. 
The main reason for this is that there is no clear policy on which object-object relationships should be stored.
First of all, it is unclear if and how  the XOC and OCEL 2.0 proposals store transitive object-object relationships  as this can cause duplication. This makes it complicated to correctly derive an object model from the stored relationships given that it is unclear which object relations are directly or indirectly connected.
Related to that is the storage of N:M relationships because storing the object-object relationships on both sides causes duplication of information.
One possible way is to reify a N:M relationship to a 1:N relationship \cite{snoeck1998existence}.
Practically, this would mean that all the object instances on N-side of the relationship store the object-instance which is on the 1-side of the relationship.
This would ensure scalability as all the object-object relationships can be traced back correctly whilst transitive relationships do not need to be stored anymore given that these can be traced back anyhow. 
This proposal however does not support schema evolution, but this could also be done with the use of dynamic attributes.
In turn, it would allow to analyze the impact a schema evolution has on the object-centric process.
A final aspect worth mentioning is that none of the event log formats allow to indicate whether an object type is a subclass of another object type.
Adding this might however provide new insights as it might be possible to discover what the typical process execution is for a superclass whilst a subclass might have a specific process execution linked to them.
Of course, it is possible to analyze this by object type but is currently not possible to know whether an object type A is a superclass of object type B.

In short, it seems that with the current object-centric event log formats, the \textbf{O2O} dimension still has the most to gain given, currently, that if an event log format supports it, scalability is not always guaranteed.
Next to that, certain intricacies of object modeling such as schema evolution, relation qualifiers and (especially) object type inheritance are not fully supported yet even though these might provide interesting insights for an object-centric process analysis.

\subsubsection{Event-Object Relations}A third insight is that both OCEL 1.0 and DOCEL require that each event has at least 1 object connected to it and each object is at least involved in one event.
Regarding the requirement that 1 event must have at least 1 object, this stems from the fact in XES every event must be assigned a case. 
Therefore a similar assumption is taken in OCEL 1.0 and DOCEL.
But there is also the possibility that an event is relevant for a process but not for a specific set of objects.
Secondly, regarding the requirement that an object must be involved in at least 1 event, it might be that redundant process information should not be stored in an object-centric event log. 
But of course, it is possible that even though certain object instances are not directly involved in an event these might be important through their relations with objects which are directly involved in the process.

\subsubsection{Final Remarks} Even though ACEL cover the most specifications, it unfortunately is not practical for analysis as it is not sufficiently normalized \cite{10.7551/mitpress/12274.003.0034}.
Next, EKG also cover a lot of specifications, but unfortunately do not have a wide support on the analysis side as is concluded in \cite{berti2023advancements} given that EKG can not directly use algorithms based on tabular data given that EKG are knowledge graphs in essence.
In third place, there is OCEL 2.0 which sees more applications from the research community but still lacks support for specific specifications such as in the \textbf{O2O} dimension.

\section{Proposed Object-Centric Event Log Refinements}\label{Object-Centric Event Log Refinement}
According to the recent object-centric overview paper \cite{berti2023advancements}, OCEL 1.0 is currently the object-centric event log format with the widest support for database extraction and object-centric process analysis.
Given that the latest OCEL 2.0 iteration included object relations and dynamic relations on top of what OCEL 1.0 supports, it is more than probable that OCEL 2.0 will be the first object-centric event log format researchers will go to for an object-centric analysis.
Therefore, we propose a set of important but feasible solutions to further improve OCEL 2.0:
\begin{itemize}
    \item \textbf{Traceable Dynamic Attributes:} Instead of using timestamps to trace attribute changes, it might be better to use foreign keys linking to both the object instance and event responsible for the change. This way, it is unambiguous which event is responsible for a change where with the timestamp it might be that two events happened simultaneously.
    \item \textbf{Scalable Object Relations:} In order to ensure scalable object relations storage, it would be beneficial, if possible, to reify all N:M object relations to 1:N relations. Hence only a limited set of object relations need to be stored whilst still being able to discover the complete object model without storing the transitive object relations which might cause duplication.
    \item \textbf{Dynamic Object Relations:} Finally, in order to support specification \textit{S18} for changing object relations, a quick solution would be to consider object relations as dynamic attributes which can also change over time. This way OCEL 2.0 would also support schema evolution.
\end{itemize}
By incorporating the previous suggestions, OCEL 2.0 would support all dimensions namely \textbf{E2E,O2O and E20} whilst keeping everything scalable and traceable. 
On top of that, it would also support two additional relevant specifications namely supporting schema evolution and relation qualifiers.

\section{Limitations and Future Work} \label{Limitations and Future Work}
Even though the framework was constructed based on insights from multiple research fields and validated by all the authors of this study, the framework would of course benefit from a wider validation from the object-centric process mining community.
Another limitation is that the conclusions of this study are based on the current insights from the investigated research fields. 
Given that, especially, the object-centric process mining field is evolving rapidly, future studies might make that certain specifications of the framework should be updated or adapted. 
In the future, investigating how an object-centric event log format can store more information whilst still remaining scalable is certainly worth investigating.
Especially, incorporating more \textbf{O2O} specifications such as schema evolution and object type inheritance whilst ensuring scalability is relevant.

\section{Conclusion} \label{Conclusion}

This study began by noting the various object-centric event log formats available for analyzing object-centric processes.
It introduces a framework for object-centric event log specifications, derived from analyzing three pertinent research areas: process mining, object-centric process modeling, and databases. Through a comparison with existing event log formats, several key findings emerge.
Firstly, it is clear that an object-centric event log format should support the tracking of attribute value changes but this is currently done in different ways each with their advantages and disadvantages.
Secondly, storing object-object relations is not an easy feat as it quickly generates scalability issues.
Related to that is that schema evolution is a well-researched topic in databases and conceptual modeling but is not always supported in object-centric event log formats (except one). 
Lastly, object type inheritance is also highlighted as important in the databases and conceptual modeling fields but is also not supported by any object-centric event log format.
With these insights, a set of feasible solutions to further improve OCEL 2.0 is proposed.
%
%
 \bibliographystyle{splncs04}
 \bibliography{Manuscript}

\end{document}